# Development of front-end readout electronics for silicon strip detectors

QIAN Yi(千奕)[1]   SU Hong (苏弘)[1]   KONG Jie（孔洁）[1,2]
DONG Cheng-Fu(董成富)[1]   MA Xiao-Li(马晓莉)[1]   LI Xiao-Gang(李小刚)[1]

[1] Institute of Modern Physics, Chinese Academy of Sciences, Lanzhou 730000, China

[2] Graduate University of Chinese Academy of Sciences, Beijing 100049, China

**Abstract:**   A front-end readout electronics system has been developed for silicon strip detectors. The system uses an application specific integrated circuit (ASIC) ATHED to realize multi-channel E&T measurement. The slow control of ASIC chips is achieved by parallel port and the timing control signals of ASIC chips are provided by the CPLD. The data acquisition is implemented with a PXI-DAQ card. The system software has a user-friendly GUI which uses LabWindows/CVI in Windows XP operating system. Test results showed that the energy resolution is about 1.22 % for alphas at 5.48 MeV and the maximum channel crosstalk of system is 4.6%. The performance of the system is very reliable and suitable for nuclear physics experiments.

**Key words:**   front-end readout electronics，system，ASIC， PXI-DAQ

**PACS:**   29.40.Mc, 29.30.Hs

## 1. Introduction

With the development of silicon strip detectors, it is widely used in nuclear physics, high energy physics, astrophysics, and so on. Over the last decade, almost all high energy physics laboratories in the world use it as a vertex detector, such as the ANKE spectrometer at the COoler SYnchrotron COSY Juelich[1], the tracking System of ALICE [2] , the tracking measurement System of GLAST [3], and so on. As for the External Target Facility of Heavy Ion Research Facility in Lanzhou (HIRFL-ETF), silicon strip detectors will be equipped in the detection system. Because of large number of channels, a highly integrated, low power electronics system is required to be developed. Based on the experimental requirements, a front-end readout electronics system has been developed by using an Application Specific Integrated Circuits (ASIC). In this system a front-end board can implement E and T measurements of multiple channels [4], and the readout is based on a PXI data acquisition (DAQ) card, which can meet the requirements of high counting rates and a large number of readout channels [5]. The minimum readout time for 48 output energy and time information is 31μs.

In this paper, Section 2 describes the readout electronics system. Section 3 introduces the DAQ system. Performance of this system is presented in Section 4. Section 5 gives a brief summary.

## 2. System overview

The system architecture is shown in Fig. 1. The readout electronics system consists of an analog board (see the dashed box in Fig. 1), a digital board, a DAQ card (PXI-6133) and a PXI chassis. To effectively reduce the crosstalk between analog and digital signals, the signals on analog board were handled by high-speed optical coupler and then were transmitted to the digital board by flat twisted-pair cables [6].

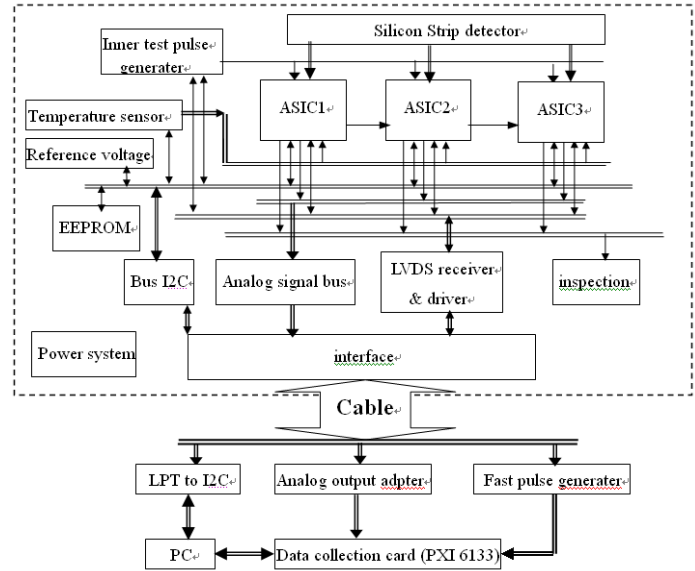

Fig. 1    Architecture of the front-end readout electronics system

## 2.1 Analog board

The analog board is composed of input circuit, ASIC chips, test signal generator, reference voltage generating circuit, LVDS (Low Voltage Differential Signal) transceivers, environmental temperature monitoring circuit, and memory cell (EEPROM). Each analog board carries three ASIC chips, and each of them can process 16 channels of energy and 16 channels of time signals. The outputs are carried on one analogue line per board and sent in a current-differential mode to the external digital board. The test signal generator on the board is a programmable pulse generator, and it can check the good functionality of all the ASIC chips and can do calibration for each channel. A single calibration capacitor is used so the injected charge is the same for all channels of three ASIC chips.

All readout and control signals from the analog board are transfered in LVDS standard to guarantee a good immunity against the electromagnetic perturbation. The trigger signal－Req, which is generated by the ASIC chips, is transferred in a point-to-point configuration in which it provides the best signal quality due to the clear path. The control signals including Start，Stop，Hold，Reset etc., are used in a multi-drop configuration. In this configuration, a driver can connect multiple receivers, so the same date can be sent to several loads.

## 2.2 Digital board

The digital board is composed of slow control, analog output adapting circuit, and fast timing pulse generating circuit.

The slow control is assured via the I2C industrial protocol. It allows in particular the configuration of the ASIC chips, together with the temperature monitoring chip, the test signal injection, the multiplexing of test channels, the board identification and the memorization of experiment parameters. And we use bus buffers (model: P82B96) to achieve the slow control cascading of more analog boards. Because the electrical characteristics of computer parallel port can meet the requirements of the I2C bus, we use it to realized I2C bus master interface for communication between PC and I2C bus devices. And the procedures are completed in Labwindows/CVI platform. The system library of LabWindows/CVI provides the functions which can be directly used to read and write the hardware port on the PC parallel port. So the hardware port on the PC parallel port is used as the input and output of SDA and SCL, and we can realize the function to send I2C star signal, send I2C stop signal, send response, send data, receive data, and etc..

Fast timing signal unit is developed based on an Altera MAXII device (model: EPM1270T144), and using Quartus to realize the software design. The main function of this circuit is to generate the necessary external fast timing pulse signal, which can make the front-end ASIC chips work normally and provides the external timing clock and trigger signals to realize data acquisition. Fig. 2 shows the time sequence diagram.

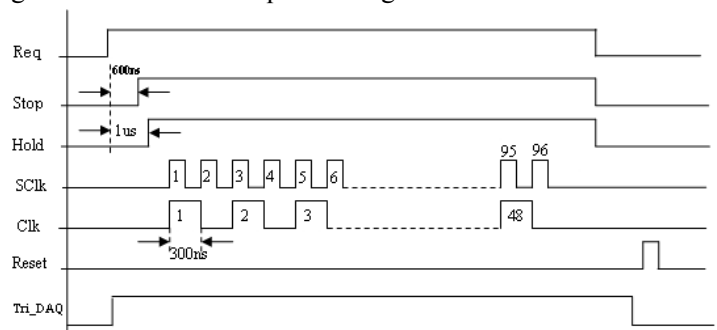

Fig. 2    Time sequence diagram generated by fast timing signal module

The analog output adapting circuit is composed of I-V transform, amplitude modulation, filtering, etc. The output differential current signal of front-end ASIC chips is translated to differential voltage signal on a 50Ω load, and this voltage signal is then amplified and converted to single-ended output by a differential receiver. The single-ended output is subsequently processed by the second-order active low-pass filter (fc=5MHz), and finally the filter's output is sent to a DAQ card via a high-speed differential driver circuit.

## 3    data acquisition system

The hardware of DAQ system is composed of a DAQ card (PXI-6133) and a PXI chassis. The DAQ software has been developed using LabWindows/CVI which makes the software design easier. A graphical user interface (GUI) allows configuring parameters of ASIC chips and other front-end on-board I2C devices, as well as the acquisition settings (physical channel, trigger source, and sampling-clock source, etc.). The acquired data can be displayed in the GUI, so that we can check whether the detectors work normally. Under the management of this program, the acquired data can be buffered into the on-board FIFO, so that the data can be read out asynchronously by the PC, and saved in files for off-line analysis. The program flow chart is shown in Fig. 3.

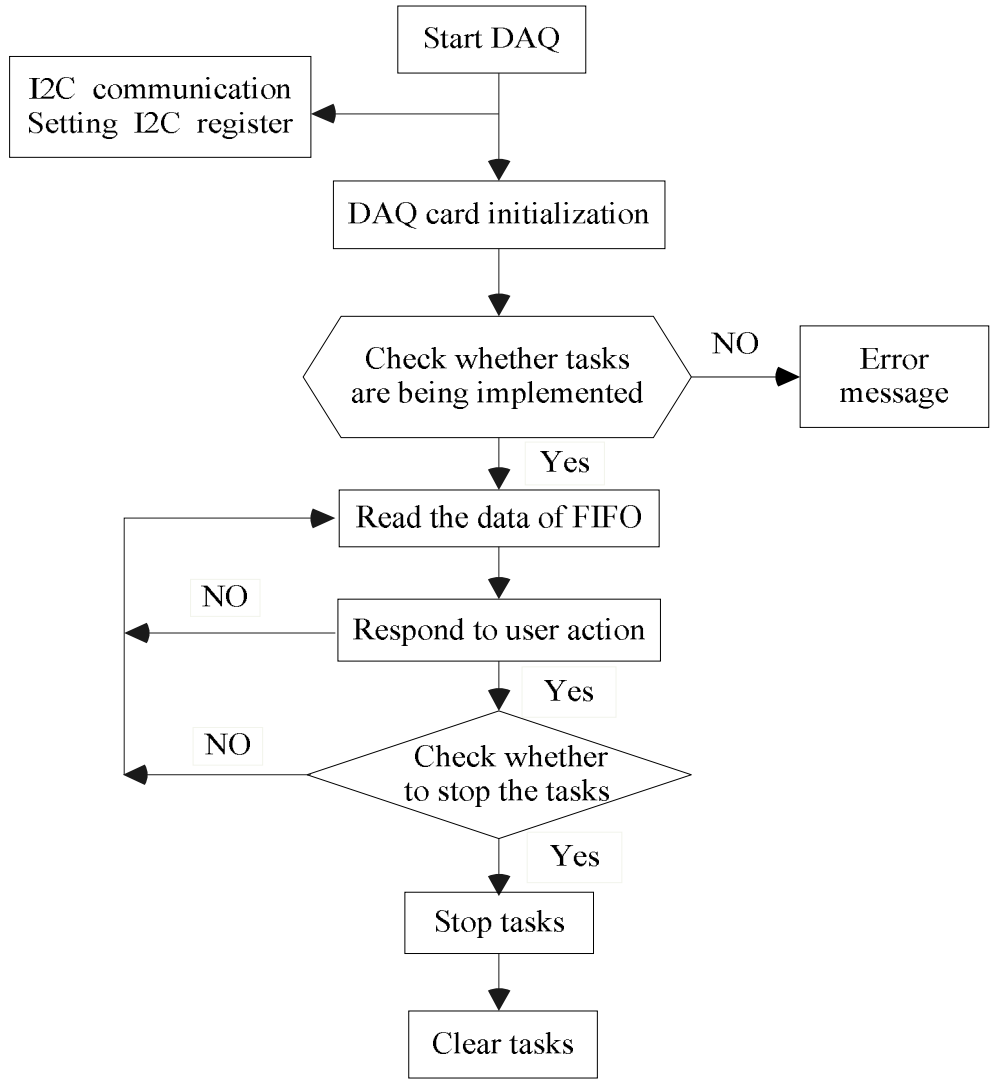

Fig. 3　Program flow chart for the DAQ software

## 4　Performance of the electronics

It shows that the system linearity is better than 0.2 % in dynamic range of 60 - 600 mV by using an accurate variable pulse generator (see Fig. 4). The crosstalk of each channel was tested, and the maximum channel crosstalk of system is 4.6 %. Fig. 5 shows the odd channel crosstalk when we input signal to dual-channel. Performance of the system has been tested with a silicon strip detector BB1 produced by Micron Semiconductor Ltd.[7]. The electronics boards are installed very close to detector and are linked to the detector via a 50 pin connector. The system works in vacuum at 7.0e-2mbar. The energy spectrum of a standard triple $\alpha$ source was measured as shown in Fig. 6. An energy resolution of ~1.2 % was achieved at 5.48 MeV

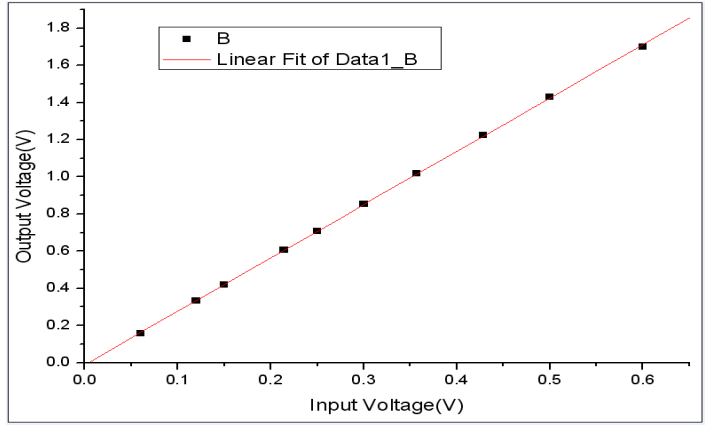

Fig. 4　　Linearity of the electronics system

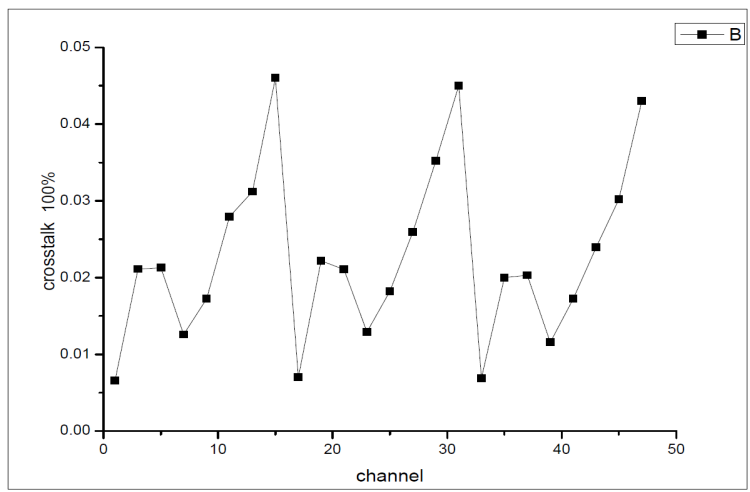

Fig. 5　　The odd channel crosstalk curve, tested by inputting signal to dual-channel

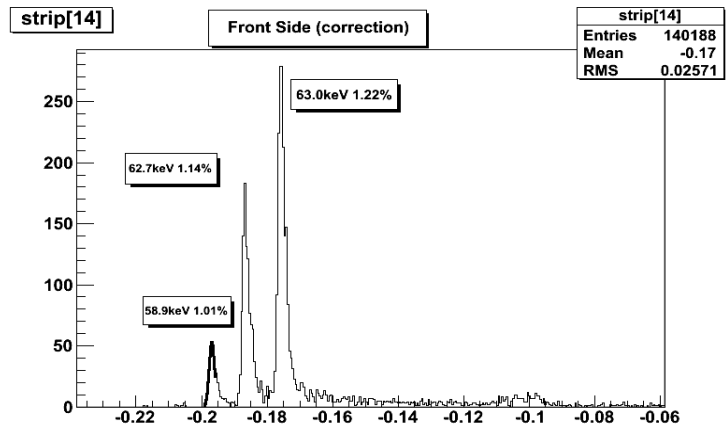

Fig. 6　　Energy spectrum of alphas source measured with a silicon strip detector.

## 5　Summary

In this paper we have described a front-end readout electronics system developed for silicon strip detectors to be equipped in the HIRFL-ETF. It shows that the system has a good energy resolution tested with a silicon strip detector and a standard alpha source. The performance of the system is accurate, reliable for future nuclear physics experiments.

**Supported by National Natural Science Foundation of China (nos.**10735060 and11005135,**) and Important Direction Project of CAS Knowledge Innovation Program (no. KJCX2-YW-N27)**

**E-mail: qianyi@impcas.ac.cn**